\documentclass[runningheads]{llncs}

\usepackage{graphicx}
\usepackage{amsmath, amssymb}
\usepackage{hhline}
\usepackage{makecell}
\usepackage{booktabs}   
\usepackage{multirow}
\usepackage{dsfont}
\usepackage{xcolor}
\usepackage{xurl}
\usepackage{subcaption}
\usepackage{hyperref}
\usepackage{float}
\usepackage{threeparttable}
\usepackage{footmisc}
\usepackage{ulem}
\newcolumntype{B}{>{\centering\arraybackslash}m{3cm}}
\newcolumntype{L}{>{\centering\arraybackslash}m{2.5cm}}

\DeclareMathOperator*{\E}{\mathop{\mathbb{E}}}

\begin{document}



\title{Redesigning Out-of-Distribution Detection on 3D Medical Images}



\author{
    Anton Vasiliuk \inst{1, 2} \and
    Daria Frolova \inst{1, 3} \and
    Mikhail Belyaev \inst{1, 3} \and
    Boris Shirokikh \inst{1, 3}
}

\institute{
    Artificial Intelligence Research Institute (AIRI), Moscow, Russia
    \and
    Moscow Institute of Physics and Technology, Moscow, Russia
    \and
    Skolkovo Institute of Science and Technology, Moscow, Russia
    \\
    \email{boris.shirokikh@skoltech.ru}
}

\authorrunning{A. Vasiliuk et al}


\maketitle
\begin{abstract}
Detecting out-of-distribution (OOD) samples for trusted \, medical image segmentation remains a significant challenge. The critical issue here is the lack of a strict definition of abnormal data, which often results in artificial problem settings without measurable clinical impact. In this paper, we redesign the OOD detection problem according to the specifics of volumetric medical imaging and related downstream tasks (e.g., segmentation). We propose using the downstream model's performance as a pseudometric between images to define abnormal samples. This approach enables us to weigh different samples based on their performance impact without an explicit ID/OOD distinction. We incorporate this weighting in a new metric called Expected Performance Drop (EPD). EPD is our core contribution to the new problem design, allowing us to rank methods based on their clinical impact. We demonstrate the effectiveness of EPD-based evaluation in 11 CT and MRI OOD detection challenges.
\end{abstract}



\section{Introduction}
\label{sec:intro}

To apply a machine learning (ML) model in clinical practice, one needs to ensure that it can be trusted when faced with new types of samples \cite{kompa2021second}. Unfortunately, the current methodology for evaluating the model's robustness does not fully address the challenges posed by volumetric medical imaging. Out-of-distribution (OOD) detection is primarily designed for the classification problem \cite{salehi2021unified}, where classification is not defined on novel classes and thus cannot be scored on the abnormal samples. Contrary, the predictions of segmentation models, the most prevalent task in medical imaging \cite{litjens2017survey}, can be scored for abnormal images. The definition of the \textit{background} class is often indistinguishable from the \textit{absence of labeled target diseases} class. So any novel occurrences can be attributed to the background, allowing us to measure segmentation quality on such samples.


Additionally, the existing OOD detection application is complicated by the continuous nature of the problem. To assess a segmentation model's reliability, two continuous aspects are addressed in a binary manner. Firstly, the segmentation quality is measured on a gradual scale, meaning that a single prediction can be partially correct rather than simply classified as either correct or incorrect. Secondly, the difference between the distribution of the data used for training the model and the distribution of novel data can also be continuous. For example, when a new location that is different from the locations in the training set is tackled as an OOD problem \cite{cao2020benchmark}, the transition is gradual rather than abrupt.

When studying these challenges using a discrete approach, where only binary classification is considered, it becomes necessary to manually select thresholds for decision-making. This approach does not allow for a proper estimation of potential errors or losses that may occur on novel data, hindering our understanding of the model's performance in such scenarios.


To address the continuous nature of novel data distributions, we can use the distance in the image space. However, the image distribution is too complex to be traceable. To overcome this challenge, we propose projecting the image distribution into a one-dimensional distribution of the model's performance scores. By doing this, any difference between the performance scores can be used as an indicator of dissimilarity in the image space. As a result, we can establish a pseudometric in the image space based on the observed variations in performance scores. Consequently, a sample is classified as abnormal if and only if it has a discernible impact on the segmentation performance.


The proposed projection provides an immediate benefit, allowing us to weigh samples based on their performance impact. We incorporate this weighting in a novel metric called Expected Performance Drop (EPD). Instead of artificial ID/OOD classification, as in previous studies, EPD measures the actual impact of an OOD detection method on the segmentation model scores. Moreover, one can train a more robust segmentation model which provides correct predictions on noisy data instead of rejecting them as abnormal. While the standard metrics indicate this case as a detection mistake, our metric reveals actually improved performance. 
So EPD explicitly addresses the question of how much performance can be maintained by applying OOD detection methods.

Overall, our contribution may be described as follows:

\begin{enumerate}
    \item We redesign the OOD detection problem into a Performance-OOD (POOD) one based on medical image segmentation specifics. POOD exploits the actual decline in the downstream performance and provides a justification for such pipeline application.
    
    
    \item We propose a new metric called EPD, which accounts for the continuous OOD nature. We evaluate the performance of the existing methods using EPD with experiments conducted in 11 OOD detection setups.
\end{enumerate}



\section{Background}
\label{sec:related}

\subsection{Problem setting}
\label{ssec:datasets:ood}

\begin{figure}[h]
    \caption{Comparison of (1-AUROC) and EPD on synthetic datasets, lower values are preferable. The data is obtained by corrupting ID data with augmentations of different severity, where severity 0 represents the original data. AUROC scores indicate that the least distorted samples are the hardest to detect, while EPD scores indicate that the most distorted samples are the most important to detect.} 
    \label{fig:severity}
    \includegraphics[width=\textwidth]{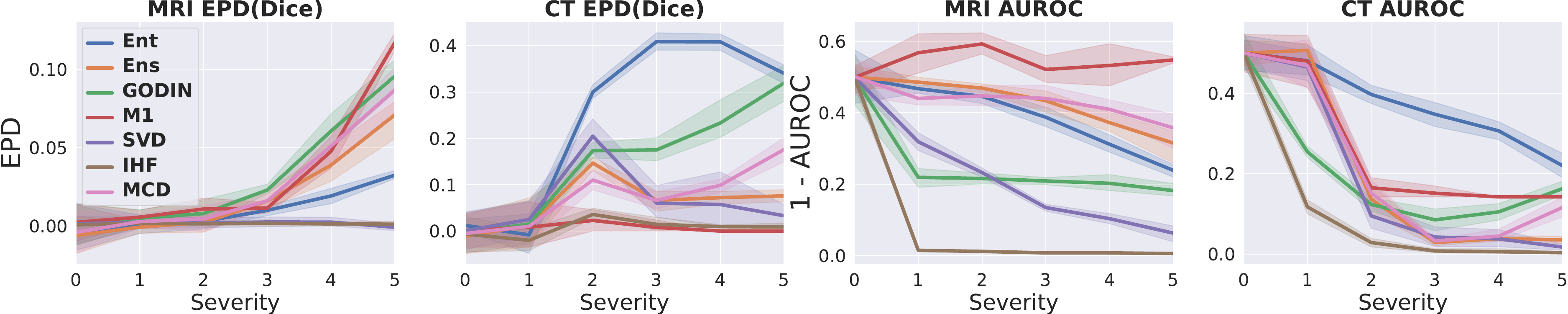}
    \centering
\end{figure}

In the field of medical imaging various anomaly sources are examined  under OOD detection framework. These anomalies can arise from continuous changes in factors such as the age of the subject \cite{karimi2022improving,mahmood2020multiscale}, image acquisition parameters \cite{lambert2022improving,karimi2022improving}, or the presence of synthesized noises  \cite{boone2022rood_bench,lambert2022improving}. Although it is crucial to detect these anomalies, they do not fit neatly into the classification-based OOD framework, as the prediction classes remain the same. While these changes could be considered within the framework of Anomaly Detection (AD), their significance is supported by a decline in the performance of downstream models. In other words, the impact of these anomaly sources becomes evident when we observe a drop in the performance of models that rely on the anomaly-free input.




Other studies have also emphasized the continuous nature of occuring anomalies. In a review of similar works \cite{salehi2021unified}, the AD definition by Grubbs (1969) is cited as finding ``samples that appear to deviate markedly from other members of the sample in which it occurs.'' As the authors note, this definition is ambiguous without a distance measure between distributions and a defined threshold. To resolve this ambiguity, we therefore suggest using a downstream model, which can induce a pseudometric in the image space. Application of the downstream model allows to assess the performance impact of the distinct anomaly samples.

Moreover, ensuring model reliability does not solely rely on rejecting certain samples. In fact, a comprehensive study was conducted by \cite{boone2022rood_bench} to investigate the methods robustness on abnormal samples. The authors' objective was to minimize the difference in segmentation performance between synthesized anomaly sources and the performance achieved on in-distribution (ID) samples. 
However, a unified framework is desired, which would enable the evaluation of model trustworthiness through both model robustness and anomaly rejection. This framework design should not be limited to specific methods and should be capable of assessing the practical impact of OOD pipelines in medical segmentation. 

The estimation of performance, considering the ability to reject samples, is closely connected to the Selective Prediction (SP) framework. SP enables us to retain a subset of data samples to gain performance on the remaining samples \cite{geifman2017selective}. However, it's important to note that the SP framework assumes the evaluation on in-distribution data. Consequently, this methodology is not suitable for estimating performance on anomalous data, making it challenging to scale its application for OOD performance estimation. 


The frameworks mentioned above have shown effectiveness in enhancing the robustness of ML models. However, they are specific to certain contexts and lack the capability to assess the potential harm caused by anomalous samples. To address this limitation, we propose a Performance-OOD detection that extends the existing methodologies. This framework aims to tackle the challenge of improving model reliability while also providing an evaluation of its practical impact. By adopting the POOD, we can generalize and expand upon the existing approaches, enabling a comprehensive assessment of model robustness as the consequences of encountering anomalous samples.

\subsection{OOD detection metrics}
\label{ssec:related:metrics}

Classification metrics are conventionally used to measure the OOD detection performance. We discuss the most commonly used ones below.

\textbf{AUROC} provides a holistic view of classifier performance across many thresholds \cite{hendrycks2016baseline}. But in practice, any algorithm works only at a specified one \cite{salehi2021unified}. We believe that the issue in applying AUROC to OOD detection is averaging scores across many \textit{irrelevant} thresholds. In OOD detection, the rate of outliers is expected to be orders of magnitude lower than the ID data rate. However, AUROC mainly scores at thresholds, where it does not preserve the majority of ID samples (e.g., $< 0.95$ true positive rate). Thus, a large part of the AUROC score is attributed to the performance in irrelevant scenarios. 

\textbf{AUPR} is another classification metric that is frequently used to assess OOD detection \cite{hendrycks2016baseline}. Besides the same issue of averaging across irrelevant thresholds, as in AUROC, the AUPR value also depends on the unknown ratio between positive and negative classes.

\textbf{FPR@TPR=$N$} is a metric that quantifies the percent of misclassified abnormal samples at a threshold where $N\%$ of ID samples is preserved \cite{bitterwolf2022breaking}. Most importantly, this metric reflects the OOD detection performance focusing on maintaining the ID data, when $N$ values are close to 100. We further build our metric upon FPR@TPR=$N$ due to its clear interpretation. Besides, one can use a robust version, FPR@TPR=$N+$, which averages over thresholds $\geq N$, or vary the value of $N$, depending on the task at hands. The proposed metric adjusts the same way FPR does when changing $N$ or $N+$. 

All described metrics use the underlying assumption of the binary nature of the OOD detection task. They a priori discard the intuitive observation that abnormal samples can impact the downstream model's performance differently. Contrary, our redesigned setup considers the varying impact of these samples.

\section{Expected Performance Drop}
\label{sec:metrics}

The impact of various anomalies on a segmentation model is not uniform. Thus, we develop a metric that assesses methods based on the downstream prediction performance. Firstly, we establish a threshold on the test ID set, aiming to retain $N\%$ of the ID data; $N=95$ by default. Threshold selection follows the motivation behind the FPR@TPR=95 metric: abnormal events are assumed to be rare, and most of the ID data should be preserved. On the occurring data, we then reject (classify as OOD) all samples above the selected threshold. We evaluate the drop in segmentation performance on the remaining data compared to the expected ID performance. Hence, achieving a zero drop is possible either through accurate OOD detection or by avoiding erroneous predictions. Mathematically, the Expected Performance Drop (EPD) metric is defined as follows:

\begin{equation}
    \label{eq:epd}
    \text{EPD} = \E_{(x, y) \sim X_{ood}} (S_0 - S(x,y)) \mathds{1} [\text{id}(x) = 1],
\end{equation}

\noindent
where $X_{ood}$ is the test OOD data, $x$ - image, $y$ - segmentation, $\text{id}(x)$ - prediction of whether $x$ is ID, $S(x,y)$ - segmentation model's score. $S_0 = \E_{(x, y) \sim X_{id}} S(x, y)$ is the expected score on the test ID set $X_{id}$. Lower EPD values are better \footnote{A reader might suspect that EPD has a failure case, a trivial detection method that labels everything as OOD ($\text{id}(x) = 0 \; \forall x$), providing EPD $= 0$. Here, we note that any method is required to retain at least 95\% TPR on the test ID set by the problem design. So the trivial detector, which outputs the same score for every image, thus the same label, is forced to label every image as ID, resulting in a valid $\text{EPD} = \E_{(x, y) \sim X_{ood}} (S_0 - S(x,y))$.}.


\paragraph{Choosing segmentation metric} The EPD metric depends on but is not restricted to any segmentation metric in particular. When dataset has nontrivial segmentation masks, we employ the Dice similarity coefficient (DSC). For instance, even though anomalies like noise or changes in acquisition protocols differ semantically, we can still acquire ground truth masks to evaluate the performance decline, as in \cite{boone2022rood_bench}. However, in scenarios where the downstream problem is absent, such as different scanning locations, the average number of false positive predictions (AvgFP) may provide more informative results.

\paragraph{Modifying segmentation model} Changing the model changes the scores $S(x, y)$ in Eq. \ref{eq:epd}. Thus, any quality improvements from methods, such as Ensemble, as well as possible losses due to the model modifications are taken into account. Alternative modification can be made independently of any OOD detection method also, aiming at improving the model's robustness. For example, one can train the same model with the extended data augmentations and potentially increase the scores $S(x, y)$ instead of improving the OOD detection method.

\section{Experiments and results}
\label{sec:exp}

\subsection{Datasets and methods}
\label{ssec:exp:materials}

Among the proposed OOD detection benchmarks on 3D medical images, we find \cite{our_paper} to be the most diverse in terms of public datasets and compared methods. The authors also link their setup to the downstream segmentation tasks. This allows us to fully re-evaluate the benchmark from the POOD perspective.

\paragraph{Datasets} Following their setup, we train lung nodules (CT) \cite{armato2011lung} and vestibular schwannoma (MRI) \cite{shapey2021segmentation} segmentation models based on 3D U-Net \cite{cciccek20163d}. The CT OOD datasets include Cancer500 \cite{cancer500}, CT-ICH \cite{hssayeni2020computed}, LiTS \cite{bilic2019liver}, Medseg9\footnote{\url{https://radiopaedia.org/articles/covid-19-3}\label{fn:medseg}}, and MIDRC \cite{tsai2021rsna}. The MRI OOD datasets are CC359 \cite{souza2018open}, CrossMoDA \cite{reuben_dorent_2022_6504722}, and EGD \cite{van2021erasmus}. In all splits, data preprocessing, and synthetic setups we follow the instructions provided in \cite{our_paper}.


\paragraph{OOD detection methods} Similarly, we explore the same set of OOD detection methods: entropy of predicted probability (Entropy) \cite{hendrycks2016baseline}, Monte-Carlo Dropout (MCD) \cite{gal2016dropout}, ensemble of models (Ensemble) \cite{lakshminarayanan2017simple},  generalized ODIN (G-ODIN) \cite{hsu2020generalized}, and singular value decomposition (SVD) \cite{karimi2022improving}. Most of them are considered either baselines or state-of-the-art in the field. We also explore two AD methods evaluated in \cite{our_paper}: intensity histogram features (IHF) and MOOD-1, an implementation of the MOOD 2022 \cite{david_zimmerer_2022_6362313} winner's solution (team CitAI).

Contrary to previous studies, operating in the POOD framework allows us to compare OOD detection methods to using segmentation model on abnormal data as is, without samples rejection. We call the latter approach \textit{no-ood}.

\paragraph{Experimental setup} Given the OOD and segmentation scores calculated following \cite{our_paper}, we compute EPD coupled with the DSC and AvgFP metrics using Eq. \ref{eq:epd}.

To address reliability enhancement through training modification, we design a second setup that differs in addition of training augmentations (random slice drop, Gaussian noise, gamma correction, and flip). As such training affect ID segmentation performance, the EPD is scored against the baseline model segmentation performance. This alternative training setup is called U-Net+augm.

\subsection{Results and discussion}
\label{ssec:exp:res}

\begin{table}[t]
    \centering
    \caption{Comparison of EPD and AUROC across the studied shifts. Methods are ranked by their mean performance. Smaller EPD (DSC) values are better.}
    \resizebox{\textwidth}{!}{%
\begin{tabular}{lcccccccc|ccccccc}
 {} & \multicolumn{8}{c}{\textbf{EPD (DSC)}} & \multicolumn{7}{c}{\textbf{AUROC}} \\

\toprule

Lung Cancer (CT) &      M1 &             SVD &            IHF &   MCD &           GODIN & Ens &             Ent & no-ood &         IHF &          SVD &        GODIN & MCD & Ens & M1 & Ent \\
\midrule
Scanner                  &  \textbf{.23}  &            .28 &            .27 &  .29 &            .29 &     .29 &            .25 &   .32 &  \textbf{.73}  &            .58 &            .72 &   .58 &        .55 &          .51 &   .65 \\
Synthetic (Elastic)       &  \textbf{.01}  &            .07 &  \textbf{.01}  &  .09 &            .18 &     .07 &            .29 &   .35 &  \textbf{.97}  &            .86 &            .85 &   .84 &        .85 &          .78 &   .65 \\
Location (Head)           &            .01 &  \textbf{.00}  &  \textbf{.00}  &  .05 &            .11 &     .10 &            .09 &   .29 &  \textbf{1.0}  &  \textbf{1.0}  &            .83 &   .85 &        .79 &          .83 &   .62 \\
Location (Liver)          &            .27 &  \textbf{.06}  &            .25 &  .42 &            .27 &     .44 &            .38 &   .47 &            .89 &  \textbf{.97}  &            .88 &   .42 &        .45 &          .61 &   .67 \\
Population (COVID-19)     &  \textbf{.26}  &            .38 &            .27 &  .29 &            .29 &     .30 &            .45 &   .50 &  \textbf{.88}  &            .74 &            .86 &   .79 &        .80 &          .66 &   .72 \\
\midrule
meanCT                    &  \textbf{.15}  &            .16 &            .16 &  .23 &            .23 &     .24 &            .29 &   .38 &  \textbf{.89}  &            .83 &            .83 &   .69 &        .69 &          .68 &   .66 \\
\midrule

\midrule
Vestibular Schwannoma (MRI) &             Ent &             SVD &            IHF &        Ens &      M1 & no-ood &           GODIN &   MCD &         IHF &          SVD &        GODIN & M1 & MCD & Ens & Ent \\
\midrule
Population (Glioblastoma) &  \textbf{-.07} &           .00 &            .00 &            .02 &            .01 &   .06 &            .14 &  .14 &  \textbf{1.0}  &  \textbf{1.0}  &            .96 &          .87 &   .44 &        .41 &   .14 \\
Population (Healthy)      &  \textbf{-.06} &            .00 &            .00 &           -.04 &            .03 &   .07 &            .00 &  .13 &  \textbf{1.0}  &  \textbf{1.0}  &  \textbf{1.0}  &          .86 &   .44 &        .16 &   .15 \\
Scanner                   &            .01 &  \textbf{.00}  &  \textbf{.00}  &            .01 &            .01 &   .03 &  \textbf{.00}  &  .02 &  \textbf{1.0}  &  \textbf{1.0}  &  \textbf{1.0}  &          .83 &   .70 &        .74 &   .59 \\
Synthetic (K-space noise) &  \textbf{.00}  &  \textbf{.00}  &  \textbf{.00}  &            .02 &            .09 &   .08 &            .06 &  .02 &  \textbf{1.0}  &            .86 &            .81 &          .24 &   .56 &        .63 &   .66 \\
Synthetic (Anisotropy)    &            .03 &  \textbf{.00}  &            .01 &            .04 &            .03 &   .06 &            .11 &  .05 &  \textbf{.98}  &            .94 &            .81 &          .57 &   .63 &        .63 &   .71 \\
Synthetic (Motion)        &            .01 &  \textbf{.00}  &  \textbf{.00}  &            .01 &            .01 &   .01 &            .06 &  .01 &  \textbf{.99}  &            .75 &            .78 &          .48 &   .57 &        .54 &   .57 \\
\midrule
meanMRI                   &  \textbf{-.01} &            .00 &            .00 &            .01 &            .03 &   .05 &            .06 &  .06 &  \textbf{1.0}  &            .93 &            .89 &          .64 &   .56 &        .52 &   .47 \\
\bottomrule
\end{tabular}
}
\label{tab:epd_roc}
\end{table}

Firstly, we compare EPD scores against the standard AUROC metric in Tab. \ref{tab:epd_roc}. EPD reflects the actual influence of the OOD detection integration into a segmentation pipeline. For lung cancer segmentation, all reviewed methods establish a considerable reliability improvement. While for vestibular schwannoma segmentation, G-ODIN performance degrades despite its high AUROC scores.

Further, the Entropy performed the poorest according to AUROC. In MRI setup this is partially due to lower OOD scores in Population shifts than on the ID set, resulting in AUROC lower 0.5. However, the EPD effectively captures the ability of the segmentation model to correctly perform under these Population shifts. This results in the lowest rejection rate, thus a better average DSC compared to the ID test set. Consequently, we demonstrate that the IHF method's AUROC score of 1.0 can be further improved if the implemented method filters out erroneous data only.

Secondly, EPD provides us with a comprehensive framework to investigate OOD detection using any relevant metric. For example, if the selection of a model is influenced by the number of false positive (FP) predictions, EPD produces a different ranking, as shown in Tab. \ref{tab:epd_n_cc}. These results highlight the effectiveness of the \textit{Ensemble}, \textit{G-ODIN}, and \textit{MCD} methods in reducing the number of FP predictions. Therefore, these methods should be preferred when minimizing FP detections is the primary criterion. None of these observations can be inferred from the AUROC metric as well as the other classification metrics.

\begin{table}[t]
    \centering
    \caption{Influence of the training pipeline on the \textit{minus} AvgFP on the CT datasets. The values are averaged across all CT shifts. Since AvgFP behaves inversely to DSC (lower is better), we negate it to preserve the same EPD relation, lower \textit{EPD (-AvgFP)} is better.}
    \resizebox{\textwidth}{!}{%
\begin{tabular}{lcccccccc|ccccccc}
 {} & \multicolumn{8}{c}{\textbf{EPD (-AvgFP)}} & \multicolumn{7}{c}{\textbf{AUROC}}  \\
\toprule
{} &         Ens &  GODIN &    MCD &    SVD &   IHF & M1 &           no-ood &    Ent &         IHF & SVD & GODIN & MCD & Ens & M1 & Ent \\
\midrule
Unet      &  \textbf{-4.71}  &  -4.57 &  -4.42 &  -3.05 &  -2.95 &       1.22 &             1.53 &   2.33 &  \textbf{.89}  &   .83 &     .83 &   .69 &        .69 &          .68 &   .66 \\
Unet+augm &            -5.19 &  -6.39 &  -5.45 &  -2.79 &  -2.94 &      -2.90 &  \textbf{-7.04}  &  -6.11 &  \textbf{.89}  &   .86 &     .77 &   .66 &        .56 &          .68 &   .61 \\
\bottomrule

\end{tabular}
}
    \label{tab:epd_n_cc}
\end{table}

Furthermore, EPD metric enables joint optimisation of the OOD methods and a downstream model's robustness. In Tab. \ref{tab:epd_n_cc}, we demonstrate how EPD advances the utilization of training augmentations to improve model reliability. Specifically, Unet+augm model without OOD rejection (no-ood) produces the lowest number of FP detection across studied methods. Additionally, further application of OOD pipelines adversely affects this performance. In contrast, the AUROC metric exhibits similar scores for both methods and cannot represent the difference of various performance quantification.



Finally, EPD excludes the criteria of whether data is abnormal. As shown in Fig. \ref{fig:severity}, AUROC gives $0.5-0.6$ score for data with minor variations. This may lead to inadequate conclusions, such as ``further research is needed to detect close-OOD samples.'' In practice, such samples do not impact segmentation performance and can be safely ignored by the OOD detection method. And EPD indicates this safe behavior with close to 0 values at Severity $\leq1$. Therefore only such samples that influence model performance are considered abnormal.

\begin{figure}[h]
    \caption{Spearman correlation values between OOD and performance scores. Correlations with p-value  $> 10^{-4}$ are indicated by 0.}
    \label{fig:correlation}
    \includegraphics[width=\textwidth]{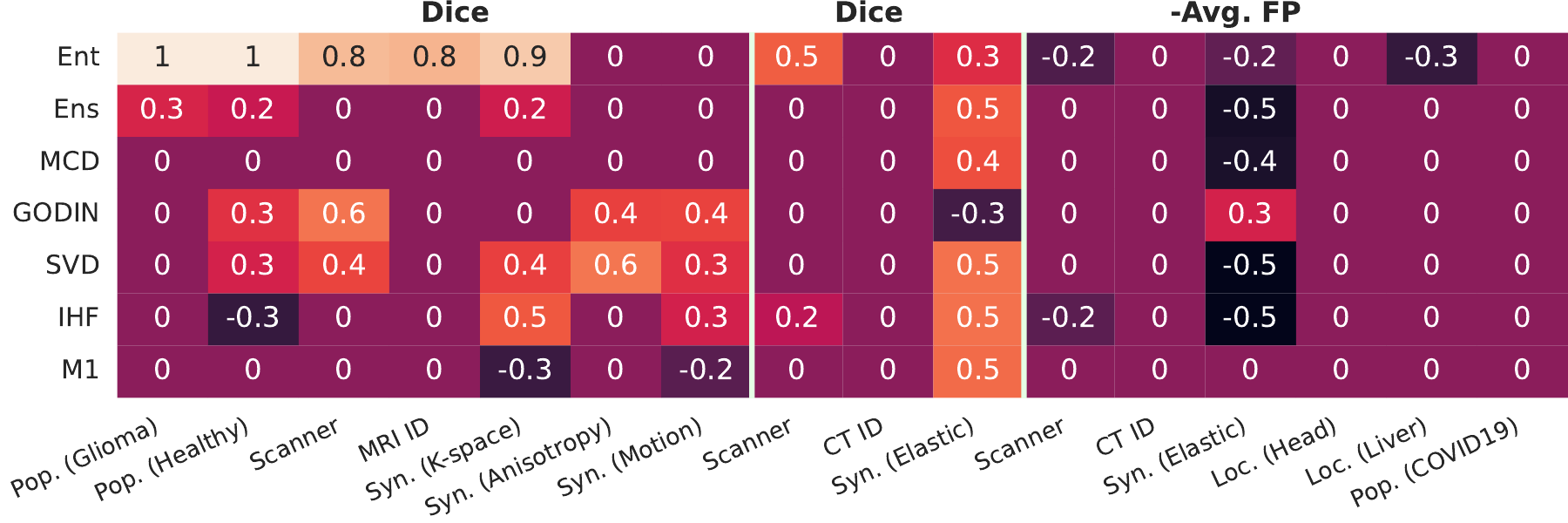}
    \centering
\end{figure}

A side benefit of the EPD metric is its ability to capture inherent correlations between the OOD score and prediction quality. As we show in Fig. \ref{fig:correlation}, such correlations exist for certain datasets and methods, thus rejecting samples affects performance on the remaining data.
EPD captures this correlation by design, resulting in accurate produced scores.


\section{Conclusion}
\label{sec:conclusion}

In this study, we reviewed the OOD detection problem, with a focus on the downstream performance drop on new data. By studying a segmentation model with the ability to reject samples, we provided a versatile perspective on the model's reliability regarding any chosen ID quality measure. Our approach enabled the analysis of arbitrary distant distributions without a requirement to define a threshold between ID and OOD. Through the application of the proposed Expected Performance Drop metric in 11 OOD detection challenges, we obtained detailed insights into the performance of segmentation models using Dice and Avg. FP scores on anomaly data. Additionally, we demonstrated that the proposed POOD framework facilitates the improvement of model reliability through both OOD pipeline implementation and robust training. Finally, with the proposed framework we evaluated the actual impact of OOD pipeline utilization, considering the potential influence on the ID segmentation performance.


\paragraph{Acknowledgments} The authors acknowledge the National Cancer Institute and the Foundation for the National Institutes of Health, and their critical role in the creation of the free publicly available LIDC/IDRI Database used in this study.

\bibliographystyle{splncs04}
\bibliography{bibliography/3_methods,bibliography/main,bibliography/1_intro}

\end{document}